\documentclass[preprint]{aastex}

\shorttitle{M22: A [Fe/H] Abundance Range Revealed}

\shortauthors{Da Costa et al}

\begin{document}

\title{M22: A [Fe/H] Abundance Range Revealed\footnote{Based on data collected at the
European Southern Observatory, Paranal, Chile, Proposal No. 77.D-0775.}}

\author{G. S. Da Costa\altaffilmark{1}, E. V. Held\altaffilmark{2}, I. Saviane\altaffilmark{3},  
and M. Gullieuszik\altaffilmark{2} }

\altaffiltext{1}{Research School of Astronomy \& Astrophysics, The Australian 
National University, Mt~Stromlo Observatory, via Cotter Rd, Weston, ACT 2611,
Australia}

\altaffiltext{2}{Osservatorio Astronomico di Padova, INAF, vicolo dell'Osservatorio 5, I-35122 Padova,
Italy}

\altaffiltext{3}{European Southern Observatory, Casilla 19001, Santiago 19, Chile}

\begin{abstract}
Intermediate resolution spectra at the Ca {\small II} triplet have been obtained for 55 candidate red
giants in the field of the globular cluster M22 with the VLT/FORS2 instrument.  Spectra were also
obtained for a number of red giants in standard globular clusters to provide a calibration of the
observed line strengths with overall abundance [Fe/H].  For the 41 M22 member stars that lie within 
the $V-V_{HB}$ bounds of the calibration, we find an abundance distribution that is substantially
broader than that expected from the observed errors alone.  We argue that this broad distribution
cannot be the result of differential reddening.  Instead we conclude that, as has long been
suspected, M22 is similar to $\omega$ Cen in having an intrinsic dispersion in heavy
element abundance.  The observed M22 abundance distribution rises sharply to a peak at 
[Fe/H] $\approx$ --1.9 with a broad tail to higher abundances: the highest abundance star in our
sample has [Fe/H] $\approx$ --1.45 dex.  If the unusual properties of $\omega$ Cen have their origin
in a scenario in which the cluster is the remnant nucleus of a disrupted dwarf galaxy, then such a
scenario likely also applies to M22.
\end{abstract}

\keywords{globular clusters: general --- globular clusters: individual (M22) --- stars: abundances}

\section{Introduction}

Beginning from observations made more that 30 years ago, we now know that the majority, if 
not all, globular clusters show star-to-star variations in the abundances of the light elements C, N, 
O, Na, Al and Mg \citep[see, for example, the reviews of][ and references therein]{RK94,GSC04}.  
The variations are in the sense that relative to `normal' stars in the cluster, carbon and oxygen are
depleted while nitrogen is enhanced, and sodium and aluminum are enhanced while magnesium 
is depleted, in the stars showing the `anomalous' abundances.  Together these effects are known as 
the O-Na anti-correlation.
Despite much work, the origin of the abundance anomalies is not well understood though the fact 
that they are seen on the main sequence in at least some globular clusters points to a process that is
intrinsic to the formation of the cluster \citep[e.g., see the discussion in][]{MGK09}.


In general, however, globular clusters are chemically homogeneous when it
comes to the abundances of the iron-peak elements, with the limits on any possible internal range
in [Fe/H] quite stringent.  For example, in their detailed analysis of high dispersion spectra of 36, 23
and 28 red giants in the clusters M5, M3 and M13, \citet{KI03} list observed standard deviations
for the cluster [Fe/H] values in the range 0.03--0.08 dex, consistent with that expected from the
errors alone.  Consequently, any intrinsic abundance spreads must be considerably smaller.  
Similarly, \citet{DA90} used their red giant branch photometry to set 3$\sigma$ upper limits of 
0.04--0.09 dex for any intrinsic heavy element abundance range in six southern clusters.

The well-established exception to the lack of [Fe/H] variations in individual clusters is the stellar
system $\omega$ Centauri, the most luminous of the Galaxy's globular clusters.  While the stars in
this cluster show the O-Na anti-correlation \citep{ND95a}, they also possess a wide range in overall 
abundance \citep[e.g.,][]{FR75, ND95b,EP02} with complex distributions of element-to-iron
abundance ratios \citep[see][ and the references therein]{Ro07}.  This indicates that $\omega$ Cen
is a system that has undergone significant chemical evolution.  Indeed the nucleosynthetic 
contributions of Type II and Type Ia supernovae are recognizable in the variations of abundance ratios with [Fe/H], as are the contributions of AGB stars \citep[e.g.,][]{ND95b, Sm00, EP02}.  All in all the 
observations point to a complex chemical history for $\omega$ Cen, with star formation likely to have 
occurred in the cluster over a period of perhaps 2 Gyr \citep[e.g.,][]{LS06}.

The differences between $\omega$ Cen and other globular clusters has led to the suggestion 
that $\omega$ Cen may have formed in a different way  --- that it is the nuclear remnant of a dwarf
galaxy which has been disrupted by the tidal field of the Milky Way \citep[e.g.,][]{KF93}. 
\citet{BF03}, for example, have shown
that such a process is dynamically plausible.  The different environment may then have facilitated
additional chemical processes that do not occur in `regular' globular clusters 
\citep[e.g.,][]{BN06,Ro07}.  
 
In this context the location of the globular cluster M54 at the center of the Sagittarius dwarf spheroidal
galaxy is particularly relevant.  The Sgr dwarf is currently being tidally disrupted by the Galaxy and in a few Gyr or less, M54 will be seen as a halo globular cluster rather than as the central star cluster of a 
dwarf galaxy \citep[e.g.,][]{DA95}.  It is therefore potentially a current day example of what may have 
been the situation for $\omega$ Cen in the distant past.  From an abundance point-of-view the situation 
is complicated because the cluster is superposed on the general Sgr field population, as well as on
that of the Sgr nucleus.  \citet{BI08} have shown that the Sgr nuclear population is metal-rich 
([Fe/H] $\approx$ --0.4) and kinematically distinct from that of M54, which has [Fe/H] $\approx$ --1.5 
dex.  Nevertheless \citet{SL95} claimed from their analysis of the width of the M54 red giant branch
in a ($I, V-I$) color-magnitude diagram that $\sigma$([Fe/H])$_{int}$ $\approx$ 0.16 dex in M54.  More 
recently \citet{BI08}
have obtained Ca {\small II} triplet spectroscopy for over 700 red giants in the central M54/Sgr
field.  They associate $\sim$425 stars with M54 and find $\sigma$([Fe/H])$_{int}$ $\approx$ 0.14, 
in good agreement with previous estimates, and an observed range in [Fe/H] from --1.8 to --1.1 dex. 
Thus the case for M54 being the second star cluster after $\omega$ Cen to possess an
internal range in [Fe/H] is strong, but confirmation from an extensive high dispersion
spectroscopic study remains to be done.  Such a study would also allow measurement of 
[element/Fe] ratios and their variation (if any) with [Fe/H], from which, as for $\omega$ Cen, 
constraints could be placed on the enrichment processes.

The other globular cluster that is often mentioned in the context of $\omega$ Cen-like abundance
variations, albeit on a smaller scale, is M22 (NGC 6656).  Unlike $\omega$ Cen and M54 which
are both very luminous clusters with M$_V$ $\approx$ --10.3 and --10.0, respectively, 
M22 is a cluster of bright but not outstanding luminosity, with M$_V$ $\approx$ --8.5 \citep{WH96}.  
It lies at low galactic latitude approximately 3 kpc from the Sun and 5 kpc from the Galactic Center 
\citep{WH96}.  Based on DDO photometry for 10 M22 red giants, \citet{HHM77} were the first to
suggest an analogy between M22 and $\omega$ Cen by noting that the observed range of ultraviolet
 excesses and CN-strengths in their M22 sample was similar to that seen in $\omega$ Cen stars 
 \citep[see also][]{HH79}.  This was followed
by the much more extensive study of \citet{NF83}, who obtained low resolution spectra for 
~$\sim$100 M22 red giants.  They reported the existence of a correlation between the strength 
of cyanogen features and the strength of the Ca {\small II} H and K lines in their M22 spectra: such a 
phenomenon was also seen in similar spectra of $\omega$ Cen red giants but was much less
marked for red giants in the `normal' cluster NGC~6752.  Based on synthetic spectra
\citet{NF83} then went on to conclude that the observed range in calcium line strengths in M22
corresponded to an abundance range $\Delta$[Ca/H] $\approx$ 0.3, and that the cluster shared the 
anomalous abundance patterns of $\omega$ Cen, though to a smaller degree.

In the ensuing decades the debate as to whether M22 does or does not share many of 
the characteristics of $\omega$ Cen, particularly as regards the presence of a range in heavy element
abundance, has swung back and forth with little consensus and a number of divergent results.  
The situation is complicated by the clear presence of differential reddening across the field of the 
cluster: $\Delta$E$(B-V)$ $\approx$ 0.06 -- 0.08 mag \citep[e.g.,][]{ATT95,MP04}.  A brief survey of
existing results \citep[see][ for additional references]{MP04} includes the following.  \citet{LBC91}
used spectra at Ca {\small II} triplet region of 10 red giants to conclude that M22 was 
similar to $\omega$ Cen in displaying Ca, Na and Fe abundance variations.  They
estimated  $\Delta$[Ca/H] $\sim$ $\Delta$[Fe/H] $\sim$ 0.4 dex, a result similar to that of 
\citet{NF83}.  On the other hand,
\citet{ATT95} concluded from Stromgren+Ca photometry of $\sim$300 red giants and
horizontal branch stars that there was no evidence for a range in [Fe/H] in the cluster.  
\cite{MP04} reached
similar conclusions from their extensive wide-field photometry of M22: the maximum metallicity spread
permitted by their data is $\Delta$[Fe/H] $\sim$ 0.1 -- 0.2 dex \citep{MP04}.  Further, \citet{II04},
using high dispersion spectra for 26 M22 red giants, were able to find a consistent set of 
spectroscopic and chemical constraints which gave acceptable stellar parameters and no requirement
for any variations in [Fe$_{\rm II}$/H].   

Most recently, \citet{MM09} analyzed high dispersion UVES spectra for 17 red giants, and lower
resolution GIRAFFE spectra for 14 stars, in the cluster.  In addition to the O-Na anti-correlation 
that is also seen in many clusters,
they identified the presence of two groups of stars whose mean abundances for the $s$-process 
elements Y, Zr and Ba differ by $\sim$0.6 dex.  The $s$-process rich group also appears to have higher
iron and calcium abundances by 0.14 $\pm$ 0.03 and 0.25 $\pm$ 0.04 dex, respectively \citep{MM09}.  
In this context it is also worth noting that \citet{Pi09}
has presented a HST-based color-magnitude diagram for M22 in which it is evident that there are
two distinct cluster sub-giant branches.  M22 thus joins other clusters such as $\omega$ Cen, 
NGC~2808 and NGC~1851 where such diagrams provide evidence for the presence of two or 
more internal cluster populations that have different properties \citep[see][ and references therein]{Pi09}. 
 
As part of a larger program to study Galactic globular clusters with uncertain abundance determinations, we have investigated 
the question of an abundance spread in M22 anew, by obtaining intermediate resolution spectra at 
the Ca {\small II} triplet of a substantial number of M22 red giant stars.  Ca {\small II} triplet
line strength measurement from such spectra is a well established technique for determining
overall abundances \citep[e.g.,][]{AD91,Ru97b,Ba08}.  In \S 2 the observations and reductions are
described, while in \S 3 a calibration of the measured line strength indices with overall abundance 
is generated.  The results for M22 are also presented in \S 3 and are discussed in \S 4.  
In brief, we find strong evidence that there is an overall abundance spread in our M22 
sample of size similar to that originally found by \citet{NF83} and \citet{LBC91}, and consistent with the
results of \citet{MM09}.  Moreover,  the M22 abundance distribution bears some similarity to that for 
$\omega$ Cen, although on a reduced scale.

\section{Observations and Reductions}

\subsection{Observations}

Short and `long' exposures of each cluster in the program were first obtained in service time with
the FORS2 instrument in imaging mode at the Cassegrain focus of the VLT-UT1 telescope,
primarily to define the targets for spectroscopic follow-up.  
The $6.8\arcmin \times 6.8\arcmin$ field-of-view was centered on the cluster for the sparser or
more distant systems, but was offset from the centers for the nearer and richer 
clusters, including
the `standard' clusters.  These latter clusters were taken from the compilation of
\cite{PVI05} and possess well established [Fe/H] values 
\citep[e.g.][]{KI03} that cover the full range of metallicities exhibited by Galactic
globular clusters.  Observation of such clusters in addition to the program clusters
allows the derivation of a line strength -- abundance calibration from the subsequent 
spectroscopic observations.  
The images were obtained in the $V$ and $I$ bands.

Point-Spread-Function photometry was carried out on both the long and short 
exposure pairs using Stetson's DAOPHOT/ALLSTAR package \citep{PS87, PS94}.  The 
resulting instrumental magnitudes were then provisionally calibrated by using color 
terms, which are small, and zero points provided by ESO as part of their routine 
quality 
control\footnote{see www.eso.org/observing/dfo/quality/FORS2/qc/photcoeff/photcoeffs\_fors2.html}.
Color-magnitude (c-m) diagrams were then
generated to allow the selection of likely cluster members as targets for the 
spectroscopic observations.  They also permit characterization of the 
principal c-m diagram features, such as horizontal-branch morphology.  We note though
that for the clusters considered here, superior photometric data are 
available from other sources, for example from \citet{MP04} for M22.  The calibration 
approach adopted means 
that the photometry is likely to be consistent with the standard $V$, $I$ system, but 
that the zero points for each cluster data set are uncertain at the 0.05--0.10 mag 
level.
However, since our analysis 
uses only {\it differential} magnitudes, such as the magnitude difference from the 
horizontal branch, 
the zero point uncertainty is not a concern.   We note that for each cluster the 
data sets from the two detectors in the camera were treated separately: stars falling on CCD1
are labeled as `1\_' followed by a running number and those falling on CCD2 as `2\_' 
(cf.\ Tables \ref{Table1},\ref{Table2}). 

The spectroscopic observations were then carried out in visitor mode with the FORS2 instrument
in MOS-mode during a two night run at the end of May in
2006.  Conditions on the first night were photometric while on the second night there was intermittent thick cloud.  On both nights the seeing was quite variable, from below 1$\arcsec$ to worse than 
2$\arcsec$.  The instrument was used with the 1028z+29 grism and the 
OG590+32 order-blocking filter.  This gives a maximum spectral coverage of $\sim$7700\AA--9500\AA\/
at a scale of 0.85\AA\/ per (binned) pixel.  The allocation of targets to slits was done in such a way 
that the wavelength interval $\sim$8200\AA--8900\AA\/  was always on the detector.  The number
of stars observed in each cluster per configuration varied from $\sim$10 to the maximum of 19,
depending on the density of likely members within the field-of-view.
The magnitude range was set as $\sim$3 mag (or somewhat less) to insure 
sufficient coverage of the red giant branch, and the exposure times were 
chosen to ensure the brightest stars were close to, but not saturated.
Two exposures were obtained to allow removal of cosmic-rays.  For the majority of clusters 
only single configurations were used, but for M22 three separate configurations were observed
to increase the sample size.  The M22 field was centered to the east of the cluster
center (see Fig.\ \ref{pos_fig}).

\subsection{Reductions}

All spectra were extracted using the {\sc fors2} pipeline version
1.2 \citep{izzolarsen08}, and in particular the
{\sc fors\_calib} and {\sc fors\_science} pipeline recipes were
used to wavelength calibrate and extract the spectra. The first recipe
takes as input the master bias frame, the screen flat-field frames
and the arc frame, and computes the wavelength calibration and distortion
map for each slit, after finding the slit positions. A catalog of
arc lines and a grism parameter table, part of the package,
are also given as input. 
The mean residual in the wavelength calibration was $0.24$ pixels, the
mean spectral resolution was $R \approx 2440$, and the mean {\sc fwhm}
of the arc lines was $3.51\pm0.07$~\AA. 

Once the calibration tables were created, they were used with the
{\sc fors\_science} recipe to reduce the target spectra. This recipe
cannot process multiple exposures simultaneously, so each frame was
reduced independently, and then the two exposures were average-combined.
The software corrects for bias and flatfield, and computes a local
sky background for each slit, to be subtracted from the object spectra.
Cosmic-ray rejection was not applied. The extraction radius was $6$~pix,
and Horne optimal extraction was applied \citep{horne86}.
The spectra were normalized by exposure time, and the wavelength
solution was aligned to a reference set of $>$20 sky lines by applying
an offset.  The median offset was $0.9$~pix. 
The final product of the pipeline is then a {\sc fits} cube containing
the fully reduced spectra.
Each object is identified by
its position and extraction window in the rectified spectrum, and
by the row number in the {\sc fits} cube holding the reduced spectra. 
To associate each spectrum to the corresponding target star, the spatial
transformation between the `rectified CCD' coordinates of the slitlets
and the original target coordinates were computed.
In this way it was easy to verify that the objects detected
in each slitlet corresponded to the expected target. 
Finally a table was created,
where for each star the J2000 coordinates, the $V$ and $I$ magnitudes,
and the name of the associated spectrum are given.

\begin{figure}
\begin{center}
\includegraphics[width=0.6\textwidth]{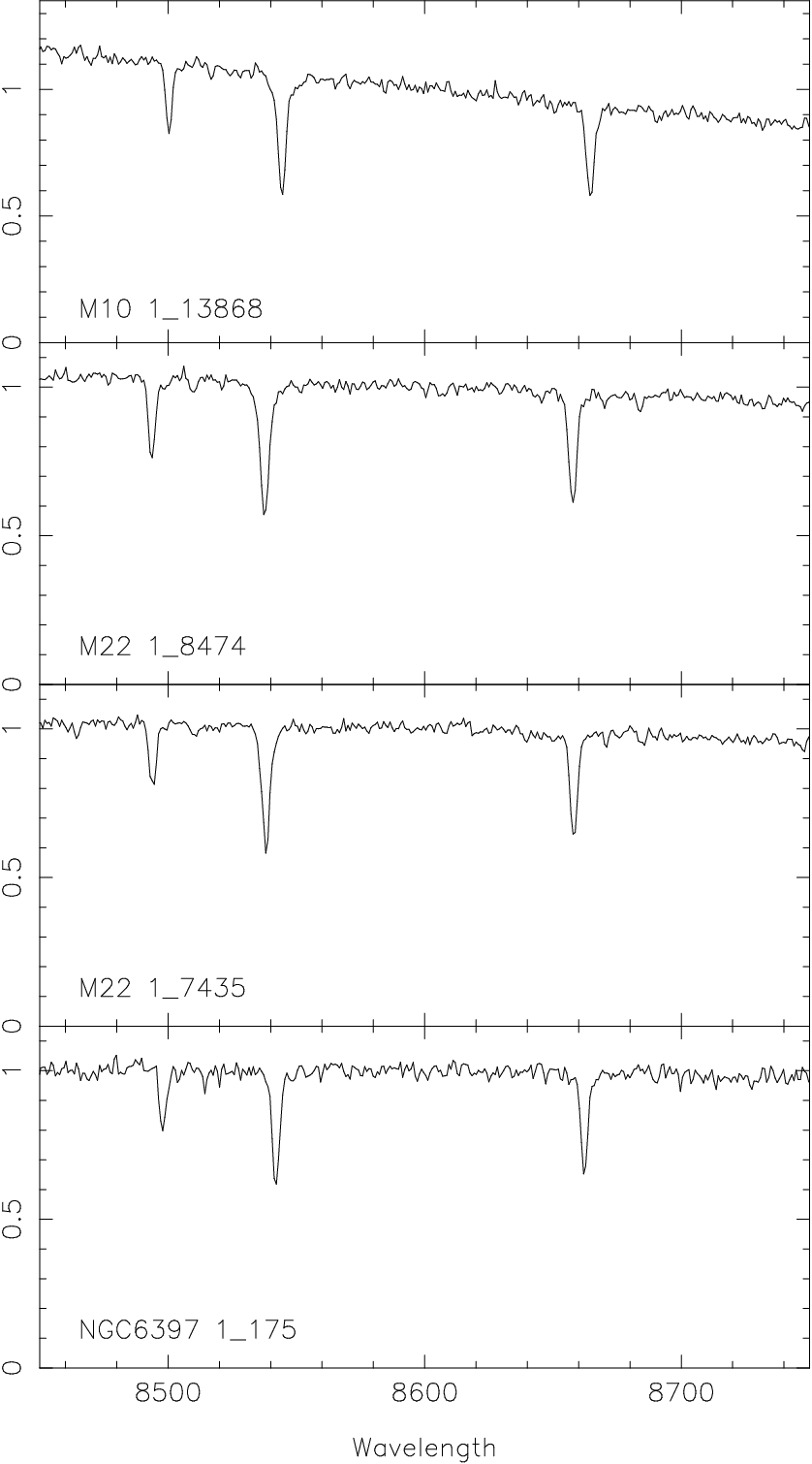}
\caption{Example spectra from the clusters NGC6397, M22 and M10.  All 4 stars have approximately
the same V--V$_{HB}$ value.  The spectra have been normalized at $\lambda$ $\approx$ 8600\AA\/
but have not otherwise been altered from those used in the analysis.
\label{spectra} }
\end{center}
\end{figure}

The S/N ratios for the final spectra varied from $\sim$110 for the brightest stars to $\sim$25 for 
the faintest in a typical exposure.  
In Fig.\ \ref{spectra} we show examples of the final spectra for a star in NGC~6397, two stars in M22,
and one star in M10.  All four stars have V--V$_{HB}$ $\approx$ --0.5 mag.

\section{Analysis}

The first step in the analysis of the reduced spectra for each cluster was measurement of
the individual radial velocities.  This was carried out with the {\sc rvidlines} task in IRAF's 
{\sc rv} package using all three of the Ca {\small II} triplet lines.  After heliocentric correction the
individual velocities were compared with the cluster radial velocity from the compilation of 
\cite{WH96}.  Obvious outliers were discarded as definite non-members and the remaining velocities 
averaged.  For the 5 standard clusters (M15, NGC 6397, M10, M4 and M71), the mean difference
between the cluster velocities determined here and that tabulated by \citet{WH96} was 
0 $\pm$ 3 km s$^{-1}$, where the uncertainty is the standard error of the mean.  

\subsection{Abundance Calibration}

For the radial velocity member stars in the standard clusters the (pseudo) equivalent widths of the 
$\lambda$8542\AA\/ and $\lambda$8662\AA\/ lines of the Ca {\small II} triplet were measured by 
fitting gaussian line profiles using the feature and continuum bands listed in \citet{AD91}.  Along with
the value, the fit also returns an uncertainty in the measured equivalent width. 
The sum of the two measurements, $W_{8542}+W_{8662}$, was then plotted against the
magnitude difference from the horizontal branch, $V-V_{HB}$,
with the measurement uncertainty in $W_{8542}+W_{8662}$ taken as the quadratic combination of 
the uncertainties for the two lines.  For all clusters
the $V$ values come directly from the photometry of the FORS2 imaging observations with 
the $V_{HB}$ values determined from the color-magnitude diagrams generated with those 
photometry sets.  As noted above, the CCD1 and CCD2  sets were treated separately though in 
practice the two $V_{HB}$ values generally agreed well.  Three M71 stars, and one NGC~6397
star, lay off the ($W_{8542}+W_{8662}$, $V-V_{HB}$) relations defined by the other cluster stars; 
we assumed these stars are non-members.  Given that both M71 and NGC~6397 have relatively
low radial velocities, the fact that velocity is an imperfect membership discriminator is not surprising.

To increase the number of standard clusters available to establish the abundance
calibration, the standard cluster observations from 
\citet{MG09} were also analysed.  These spectra were obtained with the identical instrument setup and
were reduced using the same methods as employed here.  The clusters observed were 
NGC~4590, NGC~4372, NGC~6397 (no stars in common with our data set), NGC~6752, M5 
and NGC~6171.  The line
strengths on these spectra were measured in an identical fashion to those for the standard 
clusters observed here.  The $V-V_{HB}$ values were taken from \citet{MG09}. 

\begin{figure}
\begin{center}
\includegraphics[angle=-90,width=0.9\textwidth]{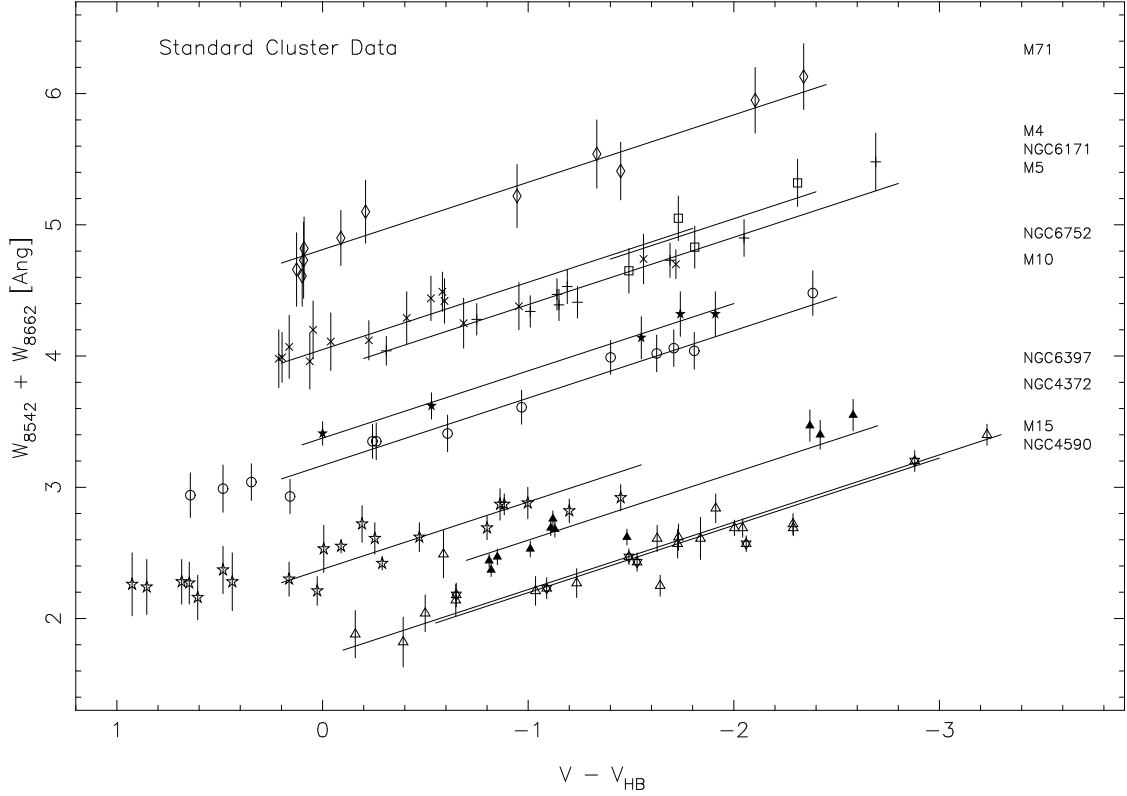}
\caption{Plot of Ca {\small II} line strength ($W_{8542}+W_{8662}$) against magnitude difference
from the horizontal branch ($V-V_{HB}$) for the 10 standard clusters.  In order of increasing 
$W_{8542}+W_{8662}$ values at $V-V_{HB}$ = --1.5, the solid lines are for clusters NGC 4590 (individual
stars plotted as open 6-pt star symbols), M15 (open triangles),
NGC 4372 (filled triangles), NGC 6397 (open 5-pt stars), M10 (open circles), NGC 6752 (filled 5-pt
stars), M5 (plus symbols), NGC 6171 (open squares), M4 (x-symbols), and M71 (open diamonds). 
The data for each cluster  has been fit with a line of slope --0.51\AA/mag for $V-V_{HB}$ $\leq$ 0.2.
Vertical bars on each point show the measurement uncertainty in the line strengths.
\label{cal_clusters_ew_fig}}
\end{center}
\end{figure}

For each of the 10 standard clusters, the ($W_{8542}+W_{8662}$, $V-V_{HB}$) data were fit with a 
straight line via least squares.  The dependence of the resulting slopes on abundance was investigated 
and found to be negligible.  A weighted average of the slopes was then formed and the resulting 
single value of  --0.51 $\pm$ 0.01 \AA/mag refitted to the data for each cluster.  
The results of this process are shown in  Fig.\ \ref{cal_clusters_ew_fig}.  Two points are worth noting.  
First, as is visible in Fig.\ 
\ref{cal_clusters_ew_fig} in the data for NGC 6397 and to a lesser extent for M10,  the slope of the 
($W_{8542}+W_{8662}$, $V-V_{HB}$) relation appears to notably flatten for $V-V_{HB}$ values 
greater than approximately +0.2 mag.  Consequently, in the 
analysis here and subsequently, only stars with $V-V_{HB}$ less than +0.2 are considered.
Second, the slope derived here is larger than the values near --0.64\AA/mag found in other studies 
\citep[e.g.,][]{AD91, Ru97a, Ba08,MG09}.  Our explanation for this difference is the following.
Because of the somewhat higher resolution of these spectra compared to those of 
\citet{AD91} for example, the gaussian-only fitting technique employed here underestimates 
the contribution to the pseudo-equivalent widths from the line profile wings, with this underestimate
being larger for stronger lines \citep[see][ for similar comments]{Ba08}.  We have investigated
this through a comparison of the $W_{8542}+W_{8662}$ values measured here with those of 
\citet{MG09}, who used a `gaussian + lorentzian' measurement technique that includes the 
contribution from the line wings.  The comparison shows an
excellent correlation over the full range of measured values for the 41 stars in 6 clusters
in common.  However, the $W_{8542}+W_{8662}$ values of \citet{MG09} are 17\% stronger, and 
when coupled with the 0.51~\AA/mag slope found here, predicts that \citet{MG09} 
should see a slope of 0.60~\AA/mag for their relation between $W_{8542}+W_{8662}$ and $V-V_{HB}$.  
\citet{MG09} actually determine a slope of 0.63 $\pm$ 0.02 \AA/mag.  Thus our explanation for 
the smaller slope found here is likely the correct one. 
However, since there is no dependence of the fitted slope on abundance, and
since the range of $V-V_{HB}$ values covered is similar for most of the clusters, our 
use of the slope determined here should not affect any of the results.  In particular, 
we note that if we consider only stars in the standard clusters with 
$V-V_{HB}$ $<$ --0.5, rather than $<$ +0.2, then the weighted mean slope found is 
--0.54 $\pm$ 0.02 \AA/mag. This value is not significantly different from that adopted 
and indicates that our choice of a limit at $V-V_{HB}$ = 0.2 for stars to include 
in the abundance calibration process does not unduly influence the results.

The variation of $W_{8542}+W_{8662}$ with $V-V_{HB}$ can then be removed by calculating
the value of the reduced equivalent width $W^\prime$, where $W^\prime$ is
the mean value of $W_{8542}+W_{8662}$+0.51($V-V_{HB}$) for the stars in each cluster, subject to
the condition that $V-V_{HB}$ $\leq$ 0.2 where necessary.  This single parameter $W^\prime$ then
needs to be calibrated with some other measure of cluster abundance.   In most situations the
calibration is made to overall abundance, designated by [Fe/H], with the [Fe/H] values chosen
from other compilations.  \citet{AD91} adopted the [Fe/H] scale of \citet{ZW84} while others, for
example \citet{Ru97b} and \citet{Ba08}, have provided calibrations to the [Fe/H] scale of 
\cite{CG97}.  \citet{KI03} have derived a new [Fe/H] scale for globular cluster abundances by
consistently analyzing high-dispersion spectra for a number of red giants in 16 key clusters 
with --2.4 $\leq$ [Fe/H] $\leq$ --0.7 dex.  In their study \citet{KI03} consider the effects of different 
model atmospheres and different color-temperature relations.  Here we adopt their [Fe/H] values
based on the MARCS model atmospheres as our fundamental calibration points. As \citet{KI03}
note, these abundances are generally about 0.2 dex lower than those of \citet{CG97}.

\begin{figure}
\begin{center}
\includegraphics[angle=-90,width=0.9\textwidth]{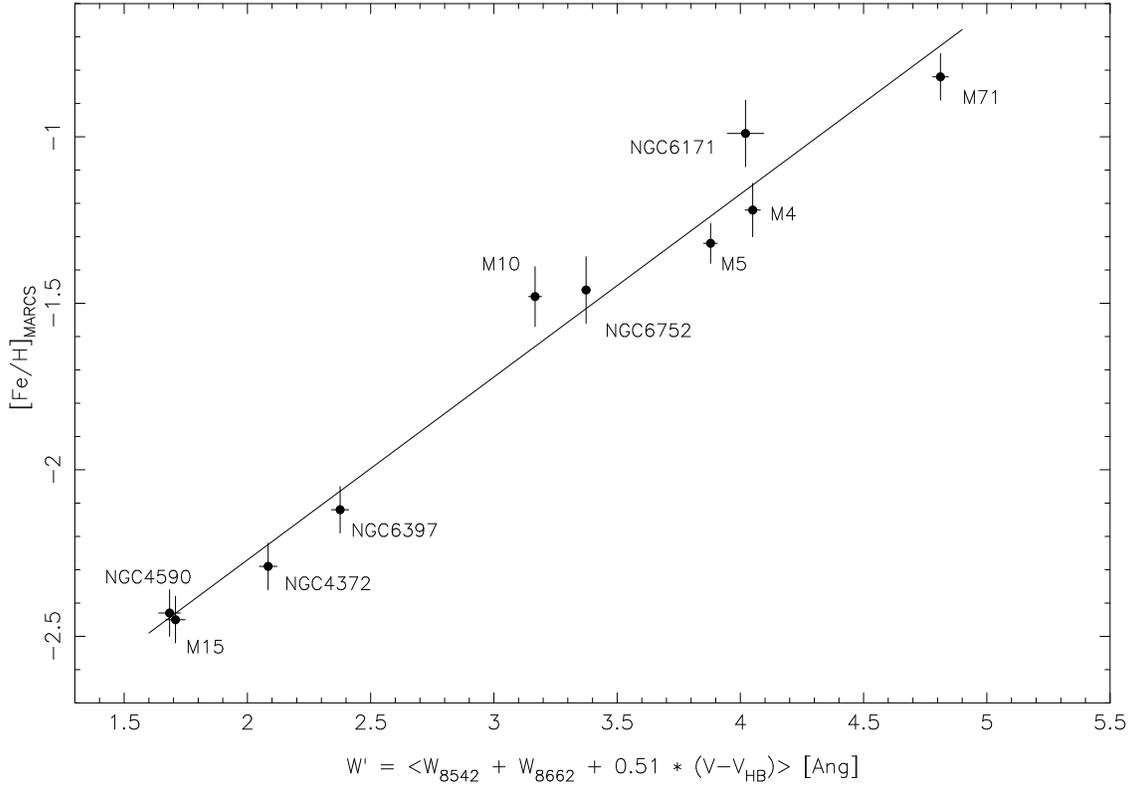}
\caption{Plot of the reduced equivalent width $W^\prime$ for the 10 standard clusters against
[Fe/H] values from \citet{KI03} derived using MARCS model atmospheres.  The fitted line has a
slope of 0.549 dex/\AA.
\label{abund_cal_fig}}
\end{center}
\end{figure}

Fig.\ \ref{abund_cal_fig} shows the relation between $W^\prime$ and [Fe/H]$_{MARCS}$ from
\citet{KI03} for our 10 standard clusters.  Over the range of [Fe/H]$_{MARCS}$ shown, the 
relation is linear with no indication of any change in slope towards the higher metallicities.  We defer 
to a subsequent paper the question of the calibration for clusters that exceed the abundance of M71,
the most metal-rich cluster included here.  The error bars for the $W^\prime$ values come from 
the dispersion of the $W_{8542}+W_{8662}$ values about the fitted slope (cf.\ Fig.\ 
\ref{cal_clusters_ew_fig}) while the errors in the [Fe/H]$_{MARCS}$ values are taken from the
discussion in the Appendix of \citet{KI03}.  A linear regression applied to the points in 
Fig.\ \ref{abund_cal_fig} yields the equation:
\begin{equation}
{\rm [Fe/H]_{MARCS}} = (0.549 \pm 0.031) \times W^\prime - (3.369 \pm 0.102)
\end{equation}
The rms dispersion about the fitted line is 0.092 dex which is consistent with the average 
uncertainty of the [Fe/H]$_{MARCS}$ values \citep{KI03}, and we adopt this relation as our
calibration to overall abundance.  To within the uncertainties our calibration relation is identical to 
that, namely:
\begin{equation}
{\rm [Fe/H]_{MARCS}} = (0.531 \pm 0.025) \times W^\prime - (3.279 \pm 0.086)
\end{equation}
 given by \citet{KI03} for the relation between [Fe/H]$_{MARCS}$
and the $W^\prime$ values of \citet{Ru97a}. 

The data for the standard cluster stars observed here are given in Table \ref{Table1}.
In successive columns we list the cluster and star ID, the RA and dec, the heliocentric
velocity, the value of $V-V_{HB}$ and the values of W$_{8542}$, W$_{8662}$ and their
sum, together with their associated uncertainties.

\begin{deluxetable}{lcccrccccrr}
\tablewidth{0pt}
\tablecaption{Standard Cluster Data \label{Table1}}
\tablecolumns{11}
\tablehead{
\colhead{ID} & \colhead{RA (2000)} & \colhead{Dec (2000)} & \colhead{Rad Vel} & \colhead{$\Delta V$}
 & \colhead{W$_{8542}$} &  \colhead{$\epsilon$} & \colhead{W$_{8662}$} & \colhead{$\epsilon$} &
\colhead{$\Sigma$W} & \colhead{$\epsilon$ } \\
& & & \colhead{(km s$^{-1}$)} & \colhead{(mag)} & \colhead{(\AA)}  &  & \colhead{(\AA)} &  & \colhead{(\AA)} &  }
\startdata
M4 1\_286   & 16 23 43.2 & --26 31 54 &   87 & --0.53 & 2.57 & 0.15 & 1.87 & 0.09 & 4.44 & 0.17 \\ 
M4 1\_1227  & 16 23 41.0 & --26 31 30 &   69 & --0.58 & 2.55 & 0.12 & 1.94 & 0.09 & 4.49 & 0.15 \\
M4 1\_1604  & 16 23 54.8 & --26 31 20 &   53 & --0.23 & 2.31 & 0.12 & 1.81 & 0.09 & 4.12 & 0.15 \\
M4 1\_3266  & 16 23 38.6 & --26 30 38 &   58 & --0.41 & 2.52 & 0.17 & 1.77 & 0.11 & 4.29 & 0.20 \\
M4 1\_4341  & 16 23 41.0 & --26 30 09 &   81 &   0.21 & 2.41 & 0.19 & 1.57 & 0.11 & 3.98 & 0.22 \\
\enddata
\tablecomments{This table is available in its entirety in a machine-readable form in the online
journal.   A portion is shown here for guidance regarding its form and content.}
\end{deluxetable}

\subsection{M22}

\begin{figure}
\begin{center}
\includegraphics[angle=-90,width=0.9\textwidth]{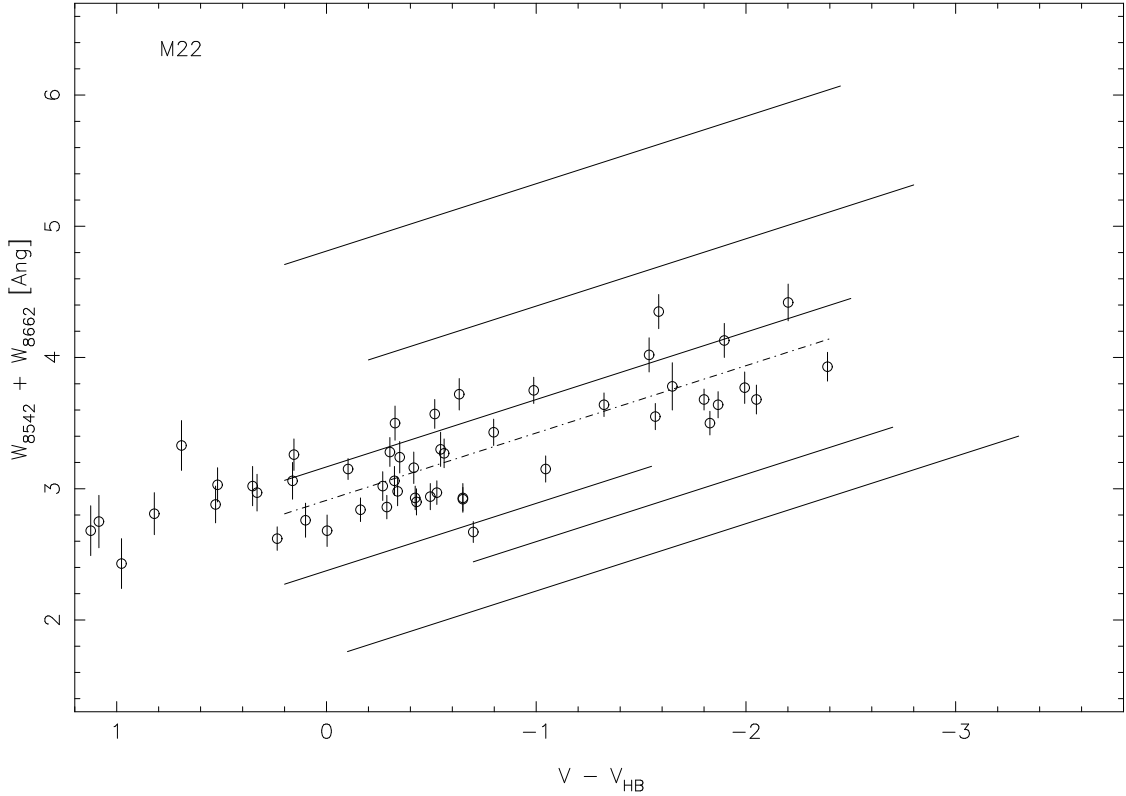}
\caption{Plot of Ca {\small II} line strength ($W_{8542}+W_{8662}$) against magnitude difference
from the horizontal branch ($V-V_{HB}$) for the 51 M22 members.  The solid lines, 
in order of increasing $W_{8542}+W_{8662}$ values at $V-V_{HB}$ = --1.5, are
the relations for the standard clusters M15 ([Fe/H]$_{MARCS}$ = --2.45),
NGC 4372 (--2.29), NGC 6397 (--2.12), M10 (--1.48), M5 (--1.32) and M71 (--0.82).  The dot-dash line
is the fit of a line of slope --0.51\AA/mag to the M22 stars with $V-V_{HB}$ $\leq$ 0.2.
Vertical bars on each point show the measurement uncertainty in the line strength.
\label{m22_ew_vhb_fig}}
\end{center}
\end{figure}

For the 55 stars in the M22 field with reduced spectra, 51 have velocities compatible with cluster
membership.  The mean velocity of these stars is --150 $\pm$ 2 km s$^{-1}$ which agrees well
with the value --148.9 $\pm$ 0.4 km s$^{-1}$ tabulated by \citet{WH96}.  The velocities of the other
4 stars clearly exclude them as members.  The $W_{8542}+W_{8662}$ values for the member stars
were then determined in the same way as for the standard cluster stars.  A plot of the values
against $V-V_{HB}$ is shown in Fig.\ \ref{m22_ew_vhb_fig}.  Shown also on the Figure is 
the best fit of a line with a slope of --0.51 \AA/mag; as for the standard clusters 
only the 41 M22 members with $V-V_{HB}$ $\leq$ 0.2 were included in the fit.  For
completeness we note that a least squares fit to these 41 points yields a slope of 
--0.49 $\pm$ 0.06 \AA/mag fully consistent with the slope derived from the standard
cluster observations.

The average $W^\prime$ value for the 41 M22 stars with $V-V_{HB}$ $\leq$ 0.2 is 
2.912 $\pm$ 0.044 (std error of the mean), which corresponds to an abundance 
[Fe/H]$_{MARCS}$ = --1.77 dex using the calibration given in
equation (1) above.  The formal uncertainty in this value is small (0.549 $\times$ 0.044 = 0.02 dex), 
but in practice the actual uncertainty will be of order of the calibration uncertainty, 
i.e.\ approximately 0.1 dex.
Comparison of this value with other determinations is not straightforward.  \citet{Ru97a} did not
observe M22 and so there is no value for this cluster given in \citet{KI03}.  \citet{CG97} give 
[Fe/H] = --1.48
$\pm$ 0.06 on their scale but \citet{KI03} note that using MARCS models, they derive abundances
that are systematically lower by about 0.2 dex than those of \citet{CG97} for the same clusters.  
\citet{ZW84} give
[Fe/H] = --1.75 $\pm$ 0.08, which is in close accord with our determination.  This is not 
surprising since \citet{KI03} note that at metallicities of this order, the difference between their
MARCS model abundances (whose scale we have adopted here) and those of \citet{ZW84} are small.  
\citet{II04} uses an abundance of [Fe/H] = --1.7 dex, presumably on the MARCS scale, while \cite{WH96}
lists [Fe/H] = --1.64 for M22.   These are again consistent with our determination.  
Similarly \citet{MM09} give a mean metallicity for M22 of [Fe/H] = --1.76 $\pm$ 0.02 (internal errors only,
weighted mean of the UVES and GIRAFFE data), essentially identical to our determination.

We give in Table \ref{Table2} the ID numbers and positions for the 51 M22
member stars, along with the measured heliocentric radial velocities in km s$^{-1}$,
the magnitude difference from the horizontal branch, $\Delta V$, the individual
$W_{8542}$ and $W_{8662}$ measurements in \AA\/ and their errors, as well as their sum,
$\Sigma$ W, and its uncertainty.  

\begin{deluxetable}{lcccrccccrr}
\tablewidth{0pt}
\tablecaption{M22 Member Data \label{Table2}}
\tablecolumns{11}
\tablehead{
\colhead{ID} & \colhead{RA (2000)} & \colhead{Dec (2000)} & \colhead{Rad Vel} & \colhead{$\Delta V$}
 & \colhead{W$_{8542}$} &  \colhead{$\epsilon$} & \colhead{W$_{8662}$} & \colhead{$\epsilon$} &
\colhead{$\Sigma$W} & \colhead{$\epsilon$ } \\
& & & \colhead{(km s$^{-1}$)} & \colhead{(mag)} & \colhead{(\AA)}  &  & \colhead{(\AA)} &  & \colhead{(\AA)} &  }
\startdata
1\_232   & 18 36 41.5 & --23 54 39 & --148 & --0.10 & 1.68 & 0.06 & 1.47 & 0.06 & 3.15 & 0.08 \\
1\_756   & 18 36 41.1 & --23 54 33 & --136 &    0.98 & 1.46 & 0.15 & 0.97 & 0.11 & 2.43 & 0.19 \\
1\_1211 & 18 36 42.3 & --23 54 28 & --144 &    0.52 & 1.73 & 0.11 & 1.30 & 0.07 & 3.03 & 0.13 \\
1\_2276 & 18 36 37.4 & --23 54 17 & --146 & --0.42 & 1.61 & 0.07 & 1.32 & 0.05 & 2.93 & 0.09 \\
1\_3666 & 18 36 36.3 & --23 54 02 & --148 & --1.57 & 2.00 & 0.09 & 1.55 & 0.05 & 3.55 & 0.10 \\

\enddata
\tablecomments{This table is available in its entirety in a machine-readable form in the online
journal.   A portion is shown here for guidance regarding its form and content.}
\end{deluxetable}

\section{Discussion}

The most striking thing about the distribution of points in Fig.\ \ref{m22_ew_vhb_fig} is the large
apparent scatter about the fitted line: the rms dispersion in $W_{8542}+W_{8662}$ at fixed
$V-V_{HB}$ for the 41 stars with $V-V_{HB}$ $\leq$ 0.2 is 0.28 \AA, which is considerably larger
than the mean measurement error, 0.11 \AA.  This stands in stark contrast to the situation for the
standard clusters.  For example, the stars observed in the standard clusters M15, NGC~6397 and M10
have mean measurement errors of 0.11, 0.11 and 0.14 \AA, respectively, values comparable to
that for M22.  Yet for these clusters the rms dispersions about the fitted line are 0.16, 0.13,
and 0.08 \AA, comparable to errors and substantially less than is the case for M22.
It is also clear from Fig.\ \ref{m22_ew_vhb_fig} that the fainter stars in M22 show a similar dispersion.  
We now investigate possible origins for this large spread.

The first possibility is that the sample is contaminated with non-members.  This seems very 
unlikely: the observed velocity dispersion of the stars classified as members is 12 km s$^{-1}$, which is 
similar to the observed dispersion for other member sets.  For example, the velocity dispersions of 
the member stars in the standard clusters M15, NGC 6397 and M10 are 20, 12 and 8 km s$^{-1}$, 
respectively.
Further, the M22 stars whose measured velocities lie furthest from the mean are not distinguished
from the remainder of the sample in Fig.\ \ref{m22_ew_vhb_fig}, nor do the stars that stand 
furthest from the mean line in the figure have velocities that are notably discrepant.  We conclude that
all the stars in Fig.\ \ref{m22_ew_vhb_fig} are indeed likely members of M22.

\begin{figure}
\begin{center}
\includegraphics[angle=-90,width=0.9\textwidth]{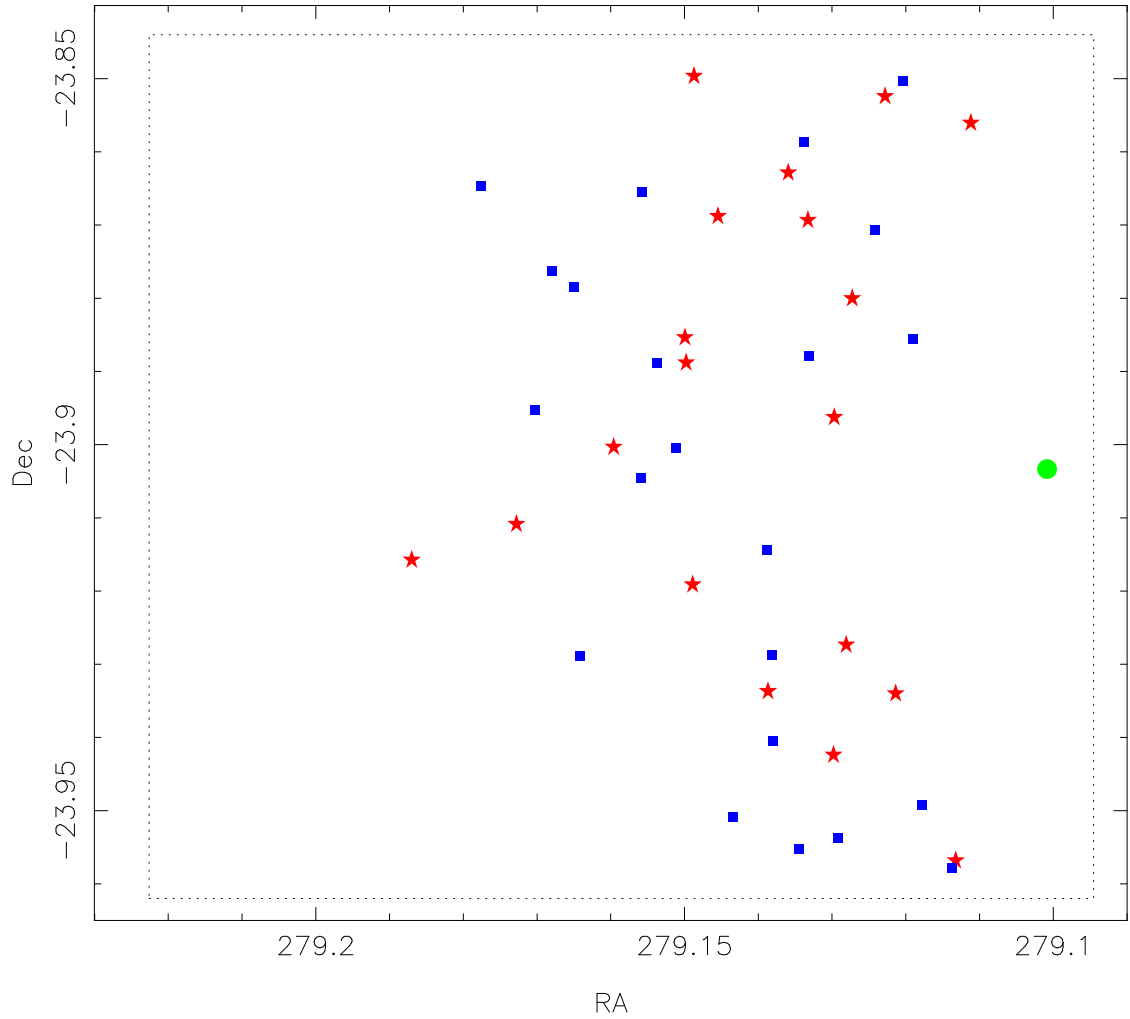}
\caption{Plot of the positions of the 19 M22 member stars that lie 
above the mean relation in Fig.\ \ref{m22_ew_vhb_fig} (red stars) and  
of the 22 stars that lie below the relation (blue squares).  The
$6.8\arcmin \times 6.8\arcmin$ field-of-view from which the observed
stars were selected is outlined by the dotted lines.  The cluster center is 
shown by the green circle.  Note that there is no obvious spatial separation 
of the red and blue symbols.
\label{pos_fig}}
\end{center}
\end{figure}

The second possibility is that the large scatter is produced by differential reddening across the
observed field.  Such reddening variations would induce scatter in the $V-V_{HB}$ values and thus
potentially increase the scatter about the mean line.  To investigate
this possibility we first examined the spatial location of the observed stars.  Specifically, we have
split the sample into two groups, those that lie above the mean line (potentially more reddened on
average, 19 stars) and those that lie below the mean line (potentially less reddened, 22 stars).  
Figure \ref{pos_fig} shows the outcome: the first group are plotted as red stars while the second 
group are plotted as blue squares.  There is clearly no straightforward segregation of the red and
blue points in the Figure.  Consequently, if reddening variations are the explanation, then the
variations must occur on small scales of order $\sim$20--30$\arcsec$, perhaps less.

While it is unclear whether reddening variations on such scales are present 
in our M22 field, we note that \citet{LB95} have shown that E($B-V$) variations
of order 0.05 mag can occur on scales as small as few arcsecs in the field of
the globular cluster M4.  Thus we need to
at least enquire into the size of the reddening variations required to produce the scatter in 
Fig.\ \ref{m22_ew_vhb_fig}.  We have done this via Monte-Carlo simulations.  We start with the
assumption that there is no intrinsic line strength spread other than that induced by the measurement
errors in the $W_{8542}+W_{8662}$ values.  The trials are conducted as follows: for each of
the 41 $V-V_{HB}$ values with $V-V_{HB}$ $\leq$ 0.2, the mean line in Fig.\ \ref{m22_ew_vhb_fig} 
is used to generate initial $W_{8542}+W_{8662}$ values.  These initial values are then altered by random draws from a gaussian distribution with mean zero and standard deviation equal to mean 
error in the observed $W_{8542}+W_{8662}$ values, which is 0.11 \AA.  In the absence of any additional broadening mechanism, the average over many trials would have a dispersion about
a fitted line of slope --0.51 \AA/pix equal to the mean observational errors, and a W$^{\prime}$ value
equal to the
observed value.  We verified that this is indeed the case.  To simulate the effects of 
differential
reddening we then perturbed the $V-V_{HB}$ values with random draws again from a gaussian, keeping the $W_{8542}+W_{8662}$ values (after perturbation with the error distribution)
unaltered.  This second gaussian has 
mean zero and a standard deviation as input, representing $\sigma($E$(B-V)$).  After the 
$V-V_{HB}$ values are perturbed, assuming A$_{V}$ = 3.2~E$(B-V)$,
a line of slope --0.51 \AA/pix is fit taking care to exclude any stars whose perturbed $V-V_{HB}$ value
exceeds 0.2, and the corresponding W$^{\prime}$ and dispersion about the line calculated.
We find that in order to consistently reproduce the observed dispersion of 0.28 \AA\/ with mean 
measurement errors of 0.11 \AA, we require $\sigma($E$(B-V)$) $\approx$ 0.12--0.15 mag. 
Clearly adopting a larger value than 3.2 for $R$ = A$_{V}$/E$(B-V)$ would increase the
dispersion about the mean line for a given $\sigma($E$(B-V)$).  However, we find that even
adopting $R$ = 3.6 \citep[cf.\ Appendix F of][]{BCP98}, the result is not materially
altered --- the minimum $\sigma($E$(B-V)$) required for the observed dispersion to
be consistently realized in the trials becomes 0.11 rather than 0.12 mag.

These values of $\sigma($E$(B-V)$) 
are considerably larger than existing estimates of the degree of differential reddening in the
field of M22.  For example, \citet{MP04} estimate a maximum {\it range} in E$(B-V)$ across the
entire face of the cluster as 0.06 mag, larger variations would be inconsistent with their data.
We conclude therefore that there is an intrinsic spread in the $W_{8542}+W_{8662}$ values
for M22, over and above that due to any differential reddening.

\subsection{The Abundance Spread in M22}

\begin{figure}
\begin{center}
\includegraphics[angle=-90,width=0.9\textwidth]{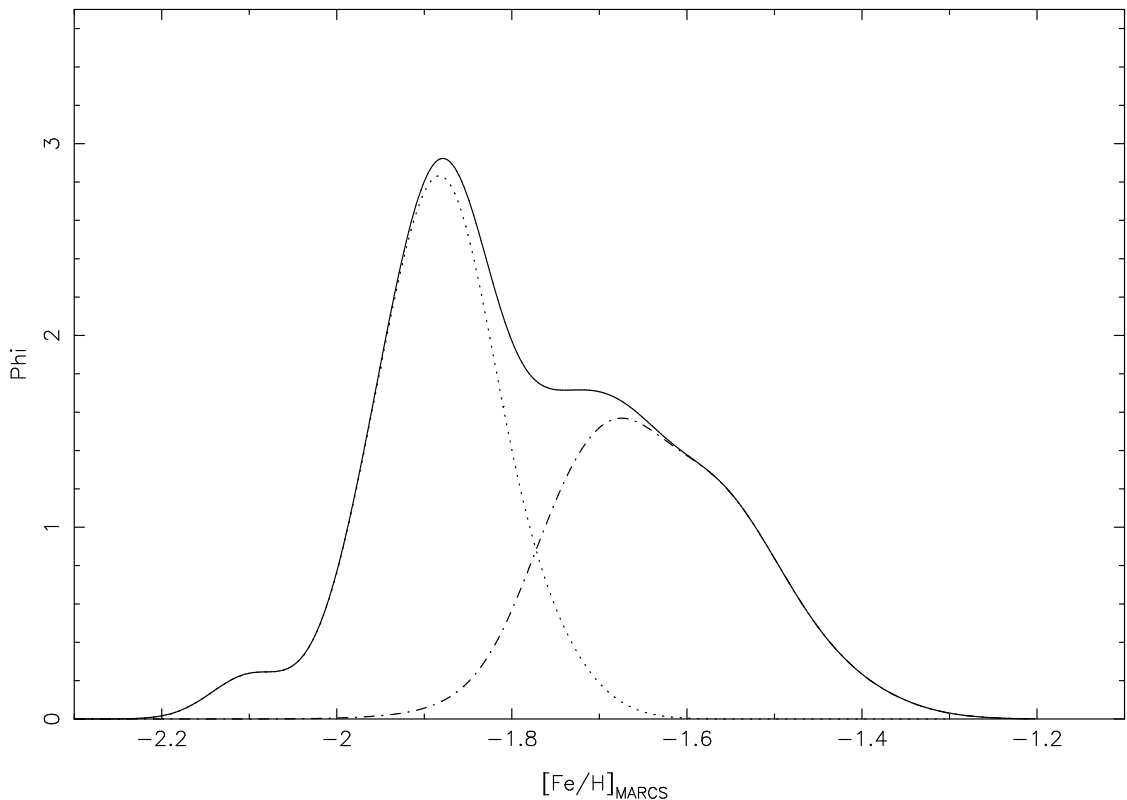}
\caption{A generalized histogram (solid line) of the individual [Fe/H]$_{MARCS}$ values for the 
41 M22 stars with $V-V_{HB}$ $\leq$ 0.2.  Each value has the abundance uncertainty from the 
uncertainty in the $W_{8542}+W_{8662}$ measurement.   Shown also are the generalized histograms
for the 22 stars below the fitted line in Fig.\ \ref{m22_ew_vhb_fig} (dotted line), and for the 19 stars
above the fitted line (dot-dash line).
\label{feh_histo_fig}}
\end{center}
\end{figure}

If the intrinsic spread in the $W_{8542}+W_{8662}$ values for the M22 stars is not due to non-member 
contamination and is not primarily the result of differential reddening, then the remaining alternative 
is that it is the consequence 
of an intrinsic abundance spread in the cluster.  Noting that for each individual star we can define a
 W$^{\prime}$ value as  $W_{8542}+W_{8662}$ + 0.51($V-V_{HB}$), equation (1) can then provide
 an abundance estimate [Fe/H]$_{MARCS}$ for each individual star.    We show in 
 Fig.\ \ref{feh_histo_fig} a 
 generalized histogram made from the individual [Fe/H]$_{MARCS}$ values and the corresponding 
 abundance uncertainties that follow from the measurement uncertainties in the $W_{8542}+W_{8662}$ 
 values.  These have a mean value of 0.06 dex.
 We also show in the figure the contributions to the total from the stars below (more metal-poor)
 and above (more metal-rich) the fitted line in Fig.\ \ref{m22_ew_vhb_fig}, which 
 corresponds to an abundance [Fe/H] = --1.77 dex.   The method assumes
 that there is no significant contribution to the abundance determinations from differential reddening.  
 This is likely to be the case for the following reason.  If the differential reddening has a range 
 $\Delta$E$(B-V)$ $\sim$ 0.06 -- 0.08 mag \citep[e.g.,][]{MP04,ATT95}, then presumably 
 $\sigma($E$(B-V)$) $\sim$ 0.02 and thus $\sigma$(A$_{V}$)
 $\sim$ 0.06 mag with the same value for  $\sigma (V-V_{HB})$.  This converts to $\sigma$([Fe/H])
 $\sim$ 0.02 dex, which is much smaller than the observed value of $\sigma$([Fe/H]) = 0.15 dex. 
 This contrasts with the photometric case where $\sigma (V-I)$ from differential reddening of size
  $\sigma($E$(B-V)$) $\approx$ 0.02 mag, and $\sigma (V-I)$ from $\sigma$([Fe/H])$_{int}$  
  $\approx$ 0.15 dex are comparable in size and thus more difficult to distinguish.
 
Figure \ref{feh_histo_fig} clearly shows that there are multiple components to the abundance
distribution in M22 --- it rises rapidly to a narrow peak at [Fe/H]$_{MARCS}$ = --1.88 yet
there is a broader tail towards higher abundances.  We can quantify this distribution in a number 
of ways.
For example, for the 22 stars in metal-poor group, the inter-quartile range is only 0.05 dex, while
for the 19 stars in the metal-rich group, the inter-quartile range is notably larger, 
0.15 dex.  The
median abundance for this group is [Fe/H]$_{MARCS}$ = --1.64 and the most metal-rich star has
[Fe/H]$_{MARCS}$ = --1.43 $\pm$ 0.07.  The most metal-poor star has [Fe/H]$_{MARCS}$ = --2.1
$\pm$ 0.05 and it is 0.14 $\pm$ 0.07 dex more metal-poor than the next most metal-poor star.  
Whether this
star represents a separate third metallicity grouping at low abundances and with a small fraction of 
the total (few percent at most), or is simply a statistical outlier, cannot be determined without a 
larger sample of member stars.  For the entire sample the inter-quartile range is 0.24 dex, which is
comparable to the M22 abundance ranges found by the earlier spectroscopic studies of 
\citet{NF83} and \citet{LBC91}, though it is clearly in conflict with the results 
of \citet{II04}.  Our results are summarised in Table \ref{Table3}.  We defer to the 
next section a comparison with the results of \citet{MM09}.  

\begin{deluxetable}{lcccc}
\tablewidth{0pt}
\tablecaption{M22 Abundance Data \label{Table3}}
\tablecolumns{5}
\tablehead{
\colhead{Sample} & \colhead{N} & \colhead{$\langle$[Fe/H]$\rangle$} & 
\colhead{$\sigma_{obs}$([Fe/H])} & \colhead{IQR} 
 }
\startdata
All              & 41 & --1.77 & 0.15 & 0.24 \\
Metal-Rich Group\tablenotemark{a} & 19 & --1.63 & 0.09 & 0.15 \\
Metal-Poor Group\tablenotemark{b} & 22 & --1.89 & 0.07 & 0.05 \\
\enddata
\tablenotetext{a}{Stars with [Fe/H] $>$ --1.77}
\tablenotetext{b}{Stars with [Fe/H] $<$ --1.77}
\end{deluxetable}

\begin{figure}
\begin{center}
\includegraphics[angle=-90,width=0.7\textwidth]{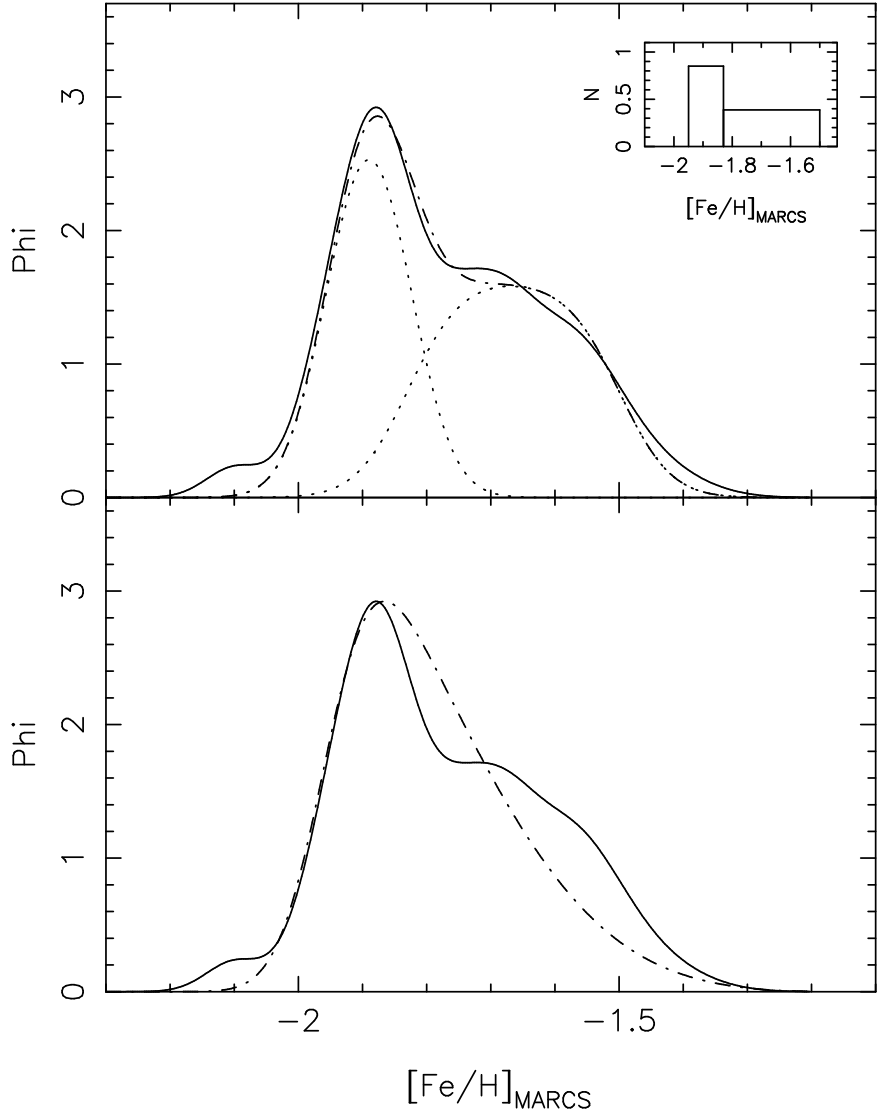}
\caption{{\it (a) Upper panel.}  The solid line is the generalized histogram 
of the individual M22 [Fe/H]$_{MARCS}$ abundances
from Fig.\ \ref{feh_histo_fig}.  The dot-dash line, which is a reasonable representation of the 
observations, is the result of convolving the abundance distribution shown in the insert with
the measurement errors.  This distribution has two components, one with abundances distributed
uniformly between [Fe/H]$_{MARCS}$ = --1.95 and --1.83, and the other distributed uniformly between
--1.83 and --1.50 dex.  The population ratio is 1 to 1.25.  The contributions of each 
component are shown as dotted lines.  
{\it (b) Lower panel.}  The solid line is again the generalized histogram of 
the individual M22 [Fe/H]$_{MARCS}$ abundances from Fig.\ \ref{feh_histo_fig}.
The dot-dash line is the abundance distribution for a
simple chemical model in which the rate of gas loss is proportional to the star formation rate.
The model assumes an initial abundance of log(Z$_{0}$/Z$_{sun}$) = --1.95 and has
log($\langle \rm{Z} \rangle$/Z$_{sun}$) = --1.77 dex corresponding to the mean abundance of the
observed sample.
\label{new_fig7}}
\end{center}
\end{figure}

We can also take `toy' models for the intrinsic abundance distribution and compare them with the 
observed distribution after convolution with the measurement errors.  One such model is shown
in the upper panel of Fig.\ \ref{new_fig7}.  Here we have assumed that the M22 
stars consist of two populations:
one with abundances uniformly distributed between [Fe/H] = --1.95 and --1.83 making up 44\% of
the total, and a second with abundances uniformly distributed between --1.83 and 
--1.50 dex 
and 56\% of the total.  This abundance distribution is shown in the insert 
in the upper panel of the figure.  Convolution with the measurement 
errors then gives the dot-dash line,
which is an acceptable representation of the observed data.  The contributions of the
two components are shown by the dotted lines.  We have not conducted an exhaustive parameter
search so it is likely that other intrinsic abundance distributions could produce similar fits to the
observations, {\it although two components with different abundance ranges does seem to be a 
requirement}. We note that the model does not reproduce the
extreme metal-poor tail of the observed distribution adequately.  A third component could no doubt 
be added to the model to fix this, but given that the metal-poor extension results from a single star, 
the addition of a third component does not seem warranted at this stage.  As noted above, a larger 
sample of member stars is needed to fully characterize the most metal-poor part of the distribution.
We note also that assuming a single abundance for the metal-poor group does not fit the
observations adequately --- a range in abundance in this population is apparently required.  However,
we have not included the effects of differential reddening, which, as outlined above, could induce
$\sigma$([Fe/H]) $\sim$ 0.02 dex.  Consequently, the extent of the metallicity range in the metal-poor
group might well be smaller than assumed in this toy model. 

We can also compare the observed abundance distribution with the predictions of simple models
of chemical evolution.  One of the simplest such model is that in which star formation proceeds
in gas whose initial abundance is Z$_{0}$ under the assumption of instantaneous recycling and with
the rate of gas loss from the system proportional to the star formation rate.  In such a model the 
metallicity distribution function f(Z) is characterized solely by Z$_{0}$ and by the mean abundance 
$\langle \rm{Z} \rangle$ \citep[e.g.,][]{FH76, NFM96}.  In the lower panel of
Fig.\ \ref{new_fig7} we show a 
comparison of the prediction of such a model with the M22 observations.  The model has been
calculated with log(Z$_{0}$/Z$_{sun}$) = --1.95 and log($\langle \rm{Z} \rangle$/Z$_{sun}$) = --1.77
dex, which is the observed sample mean abundance.  We assume Z$_{sun}$ = 0.017.  The model
distribution was then convolved with the 0.06 dex mean [Fe/H] error.  The fit to the observations is
adequate, but it does seem likely that to improve the fit, a second component with higher values for
Z$_{0}$ and $\langle \rm{Z} \rangle$ would be required \citep[cf.][]{NFM96}.  However, it is not
clear what additional insight would result from computing such a two component model given
that it does not help understand how such a model might arise physically. 

\begin{figure}
\begin{center}
\includegraphics[angle=-90,width=0.9\textwidth]{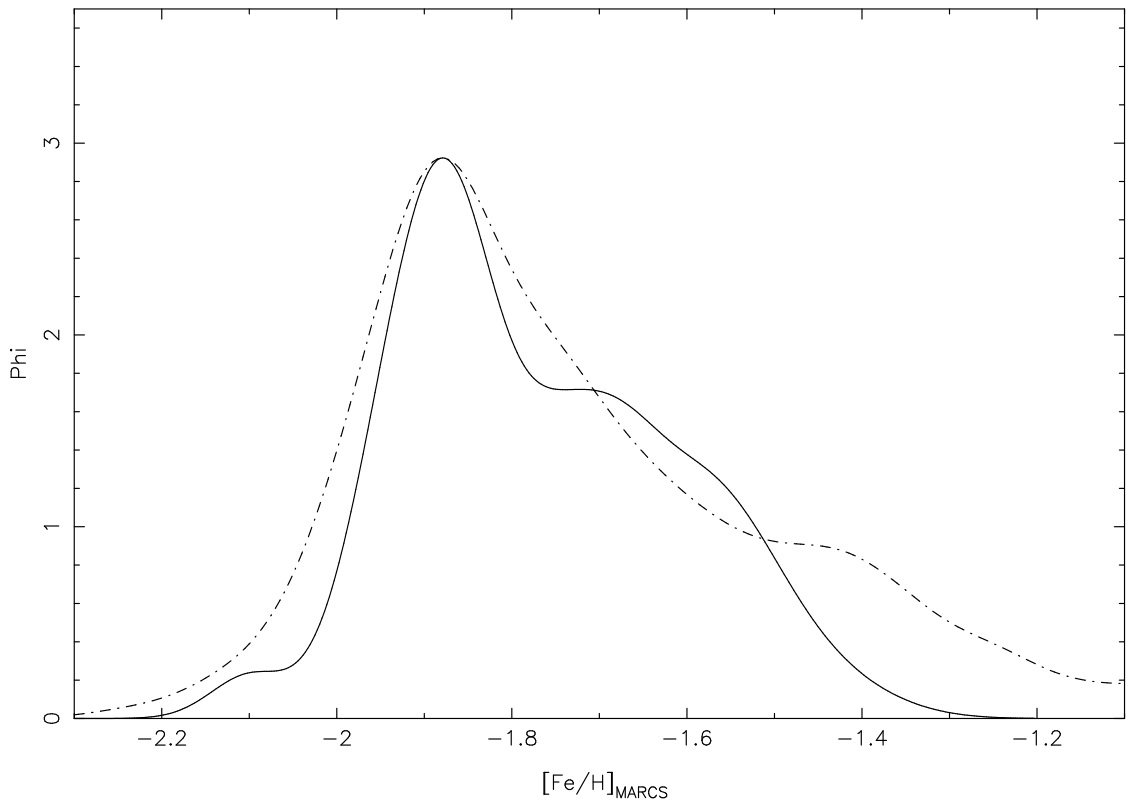}
\caption{The solid line is the generalized histogram of the individual M22 [Fe/H]$_{MARCS}$ abundances from Fig.\ \ref{feh_histo_fig}.  The dot-dash line is the abundance distribution generalised
histogram for $\omega$ Cen from \citet{NFM96}.  It has been shifted horizontally
and scaled vertically to the peak of the M22 distribution.   \label{m22_wCen}}
\end{center}
\end{figure}

Additional insight can be gained by comparing the M22 abundance distribution with that for $\omega$
Cen \citep[e.g.,][]{NFM96}.  To do this we have taken the [Ca/H] 
abundances for $\sim$500 $\omega$~Cen red giants from \cite{NFM96} and generated a generalized 
histogram using the average abundance error given by \citet{NFM96}, which, at 0.05 dex, is very similar to that for the M22 observations.  The $\omega$ Cen histogram was then shifted 
first by --0.4 dex, which corresponds to the mean [Ca/Fe] ratio for $\omega$ Cen red giants
\citep[e.g.,][]{ND95b}, and then by a further --0.09 dex to match the [Fe/H] of the peak of the M22
abundance distribution.  It was then scaled so that the peak height 
coincided with that for M22.  The outcome is shown in Fig.\ \ref{m22_wCen}.  The two distributions
are not significantly different on the metal-poor side of the peak given the smaller M22 sample.
It is clear, however, that the $\omega$ Cen distribution is broader on the metal-rich side, though 
the `rate of decline' away from the peak is similar.  It also appears that the `second abundance peak' in 
M22 is closer in abundance to the main peak than is the case for $\omega$ Cen. 
Furthermore, the $\omega$ Cen abundance distribution continues to significantly higher abundances than M22, a result that is not obviously an outcome of the smaller size of the M22 observed sample.  
Despite these differences, the general similarity between the two distributions suggests that they
could readily be the result of the same physical process or
processes, except that it has gone on longer in $\omega$ Cen, or terminated sooner in
M22, perhaps because of the difference in mass between the systems. 

Given that, in contrast to M22,  most globular clusters are homogeneous with respect to heavy element 
abundances, the similarity between the M22 and $\omega$ Cen abundance 
distributions seen in Fig.\ \ref{m22_wCen} naturally leads to the following speculation. 
The unusual properties of $\omega$ Cen have frequently led to the interpretation that the stellar
system is the nuclear remnant of a now disrupted dwarf galaxy \citep[e.g.,][]{KF93}.  Consequently,
it is plausible to suggest that M22 may also be a nuclear remnant of a disrupted dwarf galaxy.  The fact
that M54, the central star cluster of the Sgr dwarf, also shows an internal abundance range lends
further support to the suggestion.  The stellar system $\omega$ Cen lies in a tightly bound retrograde
orbit which never rises very far above the Galactic Plane \citep[e.g.,][]{DG99}.  Yet \citet{BF03} have shown that no inconsistencies arise in evolving a nucleated dwarf galaxy
in the tidal gravitational field of the Galaxy from an initial large galactocentric distance to a
nuclear remnant with an orbit resembling that of the present-day $\omega$~Cen.  As regards M22,
the compilation of \citet{DG99} shows that its orbit is fairly typical for inner halo objects \citep[cf.][]{CB07}.  It is strongly prograde ($\Theta$ = 178 $\pm$ 20 km s$^{-1}$) with apo- and pericentric 
distances of approximately 9.5 and 2.9 kpc, respectively, and a maximum height above the 
plane of 1.8 kpc \citep{DG99}.   Given the results of \citet{BF03} for $\omega$ Cen, it seems likely that
one could also start with a nucleated dwarf galaxy at large galactocentric distances and evolve it
such that the nuclear remnant has an orbit similar to that of M22.

It is also noteworthy that both $\omega$ Cen and M22  have strong blue horizontal branch 
populations, as does M54.  In the case of M22 this is not unusual given its relatively low mean 
metal abundance and relatively small distance from the Galactic Center: in the terminology 
of \citet{RZ93} and \citet{LDZ94} M22 is an `old-halo' cluster.
But it is also interesting to note that M22, $\omega$ Cen and M54, the three clusters for
which there is definite evidence for the presence of internal abundance ranges, are all in the `EHB'
classification of \citet{LGC07}.  \citet{LGC07} have shown that the EHB clusters form a kinematically
distinct group, and argue that such clusters have their origin as the cores or central star
clusters of stellar systems accreted during the earliest phases of the formation of the Galaxy.  
This is then consistent with our interpretation of M22 and its unusual properties.

\subsection{Comparison with the results of \citet{MM09}}

In a recent paper \citet{MM09} have analyzed high dispersion UVES spectra for 17 red giants in M22,
supplemented by lower resolution GIRAFFE spectra for a further 14 stars (one of which is in common
with the UVES set).  The results can be summarized as follows.  The distribution of the abundances of 
the $s$-process elements Y, Zr and Ba in the UVES sample is apparently bimodal, with one group of 
stars showing significantly larger average values of [(Y, Zr and Ba)/Fe] compared to the other.  Specifically, in the UVES sample, 7 of the 17 stars ($\sim$40\%) have 
$\langle$[$s$/Fe]$\rangle$ = 0.46 $\pm$ 0.03 while the remainder have $\langle$[$s$/Fe]$\rangle$ = 
0.02 $\pm$ 0.02 dex (errors are internal only).  The [Ba/Fe] values for the stars observed with
GIRAFFE are consistent with this result.  Moreover, for the UVES sample, the $s$-process rich group 
has both higher average [Fe/H] and [Ca/H] abundances, by 0.14 $\pm$ 0.03 dex and 0.25 $\pm$ 0.04 dex, respectively.  For the GIRAFFE sample the difference in the average [Fe/H] values is 0.18 $\pm$
0.04 dex.  The interquartile range in [Fe/H] for the full sample is 0.16 dex, with the most metal-poor
star at [Fe/H] = --1.94 and the most metal-rich at [Fe/H] = --1.59, with (internal) uncertainties of 0.09 
dex.    These values are reminiscent of the earlier work of \citet{NF83} and \citet{LBC91}, and are 
entirely consistent with the results presented here.

There are 3 stars in common between the \citet{MM09} UVES observations and the sample presented
here, with \citet{MM09} stars 200068, 200101 and 200083 corresponding to our
stars 1\_16051, 1\_3683 and 1\_10101.
For these stars, which have [Fe/H] values of --1.84, --1.74 and --1.63 in \citet{MM09}, the 
differences in [Fe/H] values in the sense of \citet{MM09} abundance minus
that derived here, are 0.09 $\pm$ 0.11, 0.02 $\pm$ 0.13 and 0.01 $\pm$ 0.11 dex, where the error
given is the combined internal uncertainty.  Clearly, although the sample is small, there is no evidence
for any systematic offset (the mean difference is 0.04 dex) and the standard deviation of the differences, 
$\pm$0.05 dex, is consistent with the errors.  Consequently, we have used a (two-sided) K-S test
to compare the [Fe/H] distribution for the 30 stars in the \citet{MM09} full sample with that for the 41 
stars with [Fe/H] values measured here.  We find that the null hypothesis, namely that the two samples 
come from the same underlying distribution, cannot be ruled out, even at a 10\% significance level.   
This has one immediate consequence.  As noted in the Introduction, \citet{Pi09} has shown that in
a {\it HST}-based c-m diagram the M22 sub-giant branch has a bimodal structure with the upper
(brighter) branch contributing 38 $\pm$ 5\% and the lower 62 $\pm$ 5\% of the total \citep[see][]{MM09}.  
One interpretation of this bimodal structure in the c-m diagram is that it reflects the abundance
distribution in the cluster, with the fainter sub-giant branch corresponding to the more metal-rich
population, with a negligible age difference.  \citet{MM09} note, however, that for a difference in 
average [Fe/H] of 0.14 dex, the predicted difference in the sub-giant branch location is $\sim$0.10 mag 
in $F606W$, smaller than the observed difference of $\sim$0.17 mag.  Further, in \citet{MM09}, the
metal-rich group is less than half the sample (37 $\pm$ 13\%), in contrast to the situation in the
c-m diagram.  Our data alleviate both of these concerns.  For example, the difference in 
average
abundance between the two populations shown in the insert  
in the upper panel of Fig.\ \ref{new_fig7} 
is 0.23 dex, which would predict a bigger magnitude difference on the sub-giant branch.  Also, the
ratio of the populations, 56\% metal-rich to 44\% metal-poor, is more in accord with the 
observed ratios in the c-m diagram.  Detailed modeling of the c-m diagram is required to place
limits on the abundance distribution, and variations in other plausible parameters such as
total CNO and age, required to reproduce the observations.

\begin{figure}
\begin{center}
\includegraphics[angle=-90,width=0.9\textwidth]{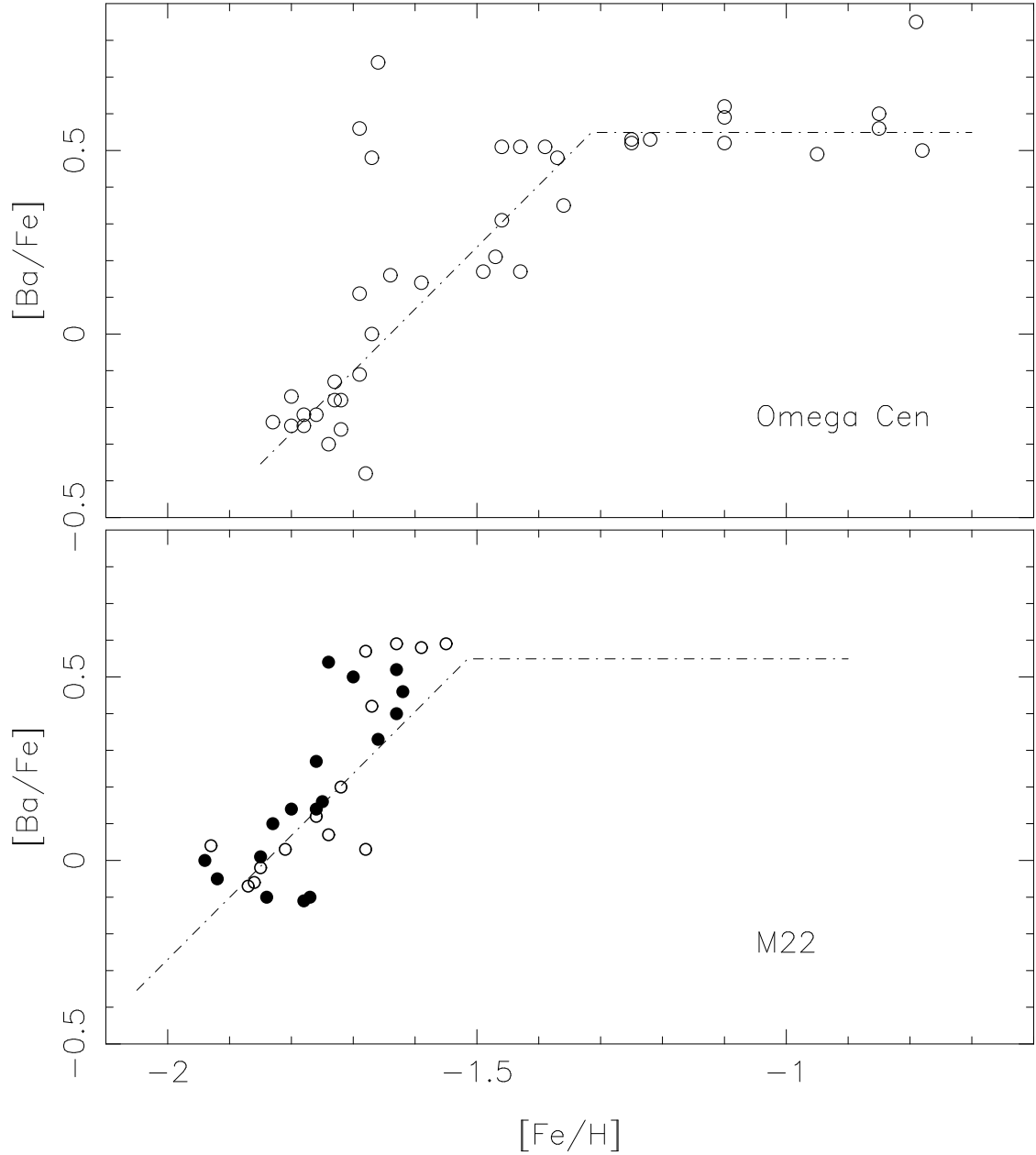}
\caption{[Ba/Fe] as a function of [Fe/H] for red giants in $\omega$ Cen using data 
from \citet{ND95b}, {\it upper
panel}, and for M22 from \citet{MM09}, {\it lower panel}.  The dot-dash lines have been fitted to the 
$\omega$ Cen distribution (see text for details) and are reproduced in the lower panel after
a shift to lower [Fe/H] values by 0.2 dex.  In the lower panel the filled circles are the stars
with UVES data, the open circles are the GIRAFFE data.  The enrichment 
processes appear similar in both stellar systems. \label{Ba_fig}}
\end{center}
\end{figure}

One further point can be made concerning M22 and $\omega$ Cen.  We argued 
above that the overall M22 abundance distribution is similar to that of $\omega$ Cen, except that 
M22 lacks the tail to higher abundances.  The availability of high dispersion abundance analyses for
M22 stars from \citet{MM09} means that the comparison can now be extended to individual
abundance ratios, using $\omega$~Cen red giant results from, for example, \citet{ND95b}.  We
show an example in Fig. \ref{Ba_fig}.  The upper panel shows [Ba/Fe] as a function of [Fe/H] for
40 $\omega$ Cen red giants from \citet{ND95b}.  The [Ba/Fe] ratio rises with increasing 
[Fe/H] until [Fe/H] $\approx$ --1.3 after which the Ba and Fe abundances change in lock-step.  
The dot-dash line is a least squares
fit to the stars with [Fe/H] between --1.9 and --1.3, excluding the three metal-poor stars with high 
[Ba/Fe], at least one of which is a carbon-star \citep[see][]{ND95b}.  
The lower panel shows the [Ba/Fe] ratios
for the 30 M22 stars in \citet{MM09}.  The dot-dash line is reproduced from the upper panel except
that it has been shifted to lower [Fe/H] by 0.2 dex.  While there are differences in detail, such as
the metal-poor stars in M22 having [Ba/Fe] $\approx$ 0.0 dex but --0.2 in $\omega$ Cen, the 
relations are clearly similar, indicating that it is likely similar enrichment processes went on in both
clusters.

\section{Conclusions}

We have shown here that there is strong evidence for the existence of a significant internal
abundance range in the globular cluster M22, consistent with the results of \citet{MM09}.   
The cluster thus joins $\omega$ Cen and 
M54 as the only Galactic globular clusters in which intrinsic [Fe/H] ranges have been established. 
The M22 abundance distribution rises sharply from low abundances to a distinct peak at [Fe/] 
$\approx$ --1.9, with a broad tail to higher abundances.  Consequently, it is probable that 
at least two components are needed to describe the distribution.  The abundance
distribution also bears a qualitative similarity to that for $\omega$ Cen, 
although the overall scale in M22 is considerably smaller.  Nevertheless, the 
similarity suggests a comparable origin for the abundance ranges in both systems.  Indeed, given
that M54 is currently the central star cluster of the Sgr dSph, and that $\omega$ Cen is frequently
postulated as being the remnant nucleus of a disrupted dwarf galaxy, it seems logical to conclude that
M22 may also be the remnant nucleus or nuclear star cluster of a disrupted dwarf galaxy. 

\acknowledgments
EVH acknowledges support from the Italian National Project PRIN MIUR 2007 ``Galactic astroarchaeology: the local route to Cosmology'' (P.I. F. Matteucci).  
We are also grateful to Anna Marino for providing details from her M22 study
and to the referee for comments that improved the presentation.

{\it Facilities:} \facility{VLT}


\begin{thebibliography}{}

\bibitem[Anthony-Twarog et al.(1995)]{ATT95} Anthony-Twarog, B. J., Twarog, B. A., 
\& Craig, J. 1995, \pasp, 107, 32

\bibitem[Armandroff \& Da~Costa(1991)]{AD91} Armandroff, T. E., \& Da Costa, G. S. 1991, \aj, 101,
1329

\bibitem[Battaglia et al.(2008)]{Ba08} Battaglia, G., Irwin, M., Tolstoy, E., Hill, V., Helmi, A., Letarte, B.,
\& Jablonka, P. 2008, \mnras, 383, 183


\bibitem[Bekki \& Freeman(2003)]{BF03} Bekki, K., \& Freeman, K. C. 2003, \mnras, 36, L11

\bibitem[Bekki \& Norris(2006)]{BN06} Bekki, K., \& Norris, J. E. 2006, \apj, 637, 109

\bibitem[Bellazzini et al.(2008)]{BI08} Bellazzini, M., Ibata, R. A., Chapman, S. C., Mackey, A. D., 
Monaco, L., Irwin, M. J., Martin, N. F., Lewis, G. F., \& Dalessandro, E. 2008, \aj, 136, 1147

\bibitem[Bessell et al.(1998)]{BCP98} Bessell, M. S., Castelli, F., \& Plez, B. 1998,
\aap, 333, 231

\bibitem[Carretta \& Gratton(1997)]{CG97} Carretta, E. \& Gratton, R. 1997, \aaps, 121, 95

\bibitem[Carollo et al.(2007)]{CB07} Carollo, D., Beers, T. C., Lee, Y. S., Chiba, M., Norris, J. E., 
Wilhelm, R., Sivarani, T., Marstellar, B., Munn, J. A., Bailer-Jones, C. A. L., Fiorentin, P. R., \&
York, D. G. 2007, \nat, 450, 1020

\bibitem[Da~Costa \& Armandroff(1990)]{DA90} Da Costa, G. S., \& Armandroff, T. E. 1990, \aj, 100, 162

\bibitem[Da~Costa \& Armandroff(1995)]{DA95} Da Costa, G. S., \& Armandroff, T. E. 1995, \aj, 109, 2533

\bibitem[Dinescu et al.(1999)]{DG99} Dinescu, D. I., Girard, T. M., \& van Altena, W. F. 1999, \aj, 117, 
1792

\bibitem[Freeman(1993)]{KF93} Freeman, K. C. 1993, The Globular Cluster---Galaxy Connection (ASP
Conf.\ Ser.\ 48), ed.\ G. H. Smith \& J. P. Brodie (San Francisco, CA: ASP), 608

\bibitem[Freeman \& Rodgers(1975)]{FR75} Freeman, K. C., \& Rodgers, A. W. 1975, \apj, 201, L71

\bibitem[Gratton et al.(2004)]{GSC04} Gratton, R., Sneden, C., \& Carretta, E. 2004, \araa, 42, 385

\bibitem[Gullieuszik et al.(2009)]{MG09} Gullieuszik, M., Held, E. V., Saviane, I., \& Rizzi, L. 2009,
\aap, 500, 735

\bibitem[Harris(1996)]{WH96} Harris, W.E. 1996, \aj, 112, 1487

\bibitem[Hartwick(1976)]{FH76} Hartwick, F. D. A. 1976, \apj, 209, 418

\bibitem[Hesser \& Harris(1979)]{HH79} Hesser, J. E., \& Harris, G. L. 1979, \apj, 234, 513

\bibitem[Hesser, Hartwick \& McClure(1977)]{HHM77} Hesser, J. E., Hartwick, F. D. A., \& McClure,
R. D. 1977, \apjs, 33, 471

\bibitem[Horne(1986)]{horne86} Horne, K. 1986, \pasp, 98, 609

\bibitem[Ivans et al.(2004)]{II04} Ivans, I. I., Sneden, C., Wallerstein, G., Kraft, R. P., Norris, J. E.,
Fulbright, J. P., \& Gonzalez, G. 2004, Mem.\ S. A. It., 75, 286

\bibitem[Izzo \& Larsen(2008)]{izzolarsen08} Izzo, C., \& Larsen, J. M., 2008, FORS Pipeline User 
Manual, VLT-MAN-ESO-19500-4106, Issue 2.0, ESO Garching, 2008-09-09 

\bibitem[Kraft(1994)]{RK94} Kraft, R. P. 1994, \pasp, 106, 553

\bibitem[Kraft \& Ivans(2003)]{KI03} Kraft, R. P., \& Ivans, I. I. 2003, \pasp, 115, 143

\bibitem[Lehnert et al.(1991)]{LBC91} Lehnert, M. D., Bell, R. A., \& Cohen , J. G. 1991, \apj, 367, 514

\bibitem[Lee et al.(1994)]{LDZ94} Lee, Y.-W., Demarque, P., \& Zinn, R. 1994, \apj, 423, 248

\bibitem[Lee et al.(2007)]{LGC07} Lee, Y.-W., Gim, H. B., \& Casetti-Dinescu, D. I. 2007, \apj, 661, L49

\bibitem[Lyons at al.(1995)]{LB95} Lyons, M. A., Bates, B., Kemp, S. N., \&
Davies, R. D. 1995, \mnras, 277, 113

\bibitem[Marcolini et al.(2009)]{MGK09} Marcolini, A., Gibson, B. K., Karakas, A. I., \& 
S\'{a}nchez-Bl\'{a}zquez, P. 2009, \mnras, 395, 719

\bibitem[Marino et al.(2009)]{MM09} Marino, A. F., Milone, A. P., Piotto, G., Villanova, S., Bedin, L. R.,
Bellini, A., \& Renzini, A. 2009, \aap, accepted for publication (arXiv:0905.4058v1)

\bibitem[Monaco et al.(2004)]{MP04} Monaco, L., Pancino, E., Ferraro, F. R., \& Bellazzini, M. 2004,
\mnras, 349, 1278


\bibitem[Norris \& Da~Costa(1995a)]{ND95a} Norris, J. E., \& Da Costa, G. S. 1995a, \apj, 441, L81

\bibitem[Norris \& Da~Costa(1995b)]{ND95b} Norris, J. E., \& Da Costa, G. S. 1995b, \apj, 447, 680

\bibitem[Norris \& Freeman(1983)]{NF83} Norris, J., \& Freeman, K. C. 1983, \apj, 266, 130

\bibitem[Norris et al.(1996)]{NFM96} Norris, J. E., Freeman, K. C., \& Mighell, K. J. 1996, \apj, 462, 241

\bibitem[Pancino et al.(2002)]{EP02} Pancino, E., Pasquini, L., Hill, V., Ferraro, F. R., \& Bellazzini, M.
2002, \apj, 568, L101

\bibitem[Piotto(2009)]{Pi09} Piotto, G. 2009, The Ages of Stars, IAU Symposium 258, ed.\ E. E.
Mamajek, D. R. Soderblom, \& R. F. G. Wyse (Cambridge: CUP), 233 (arXiv:0902.1422v1)




\bibitem[Romano et al.(2007)]{Ro07} Romano, D., Matteucci, F., Tosi, M., Pancino, E., Bellazzini, M.,
Ferraro, F. R., Limongi, M., \& Sollima, A. 2007, \mnras, 376, 405

\bibitem[Pritzl et al.(2005)]{PVI05} Pritzl, B. J., Venn, K., \& Irwin M. 2005. \aj, 130, 2140

\bibitem[Rutledge et al.(1997a)]{Ru97a} Rutledge, G. A., Hesser, J. E., Stetson, P. B., Mateo, M.,
Simard, L., Bolte, M., Friel, E. D., \& Copin, Y. 1997, \pasp, 109, 883

\bibitem[Rutledge et al.(1997b)]{Ru97b} Rutledge, G. A., Hesser, J. E., \& Stetson, P. B. 1997, \pasp, 109,
907

\bibitem[Sarajedini \& Layden(1995)]{SL95} Sarajedini, A., \& Layden, A. C. 1995, \aj, 109, 1086

\bibitem[Smith et al.(2000)]{Sm00} Smith, V. V., Suntzeff, N. B., Cunha, K., Gallino, R., Busso, M.,
Lambert, D. L., \& Straniero, O. 2000, \aj, 119, 1239

\bibitem[Stanford et al.(2006)]{LS06} Stanford, L. M., Da Costa, G. S., Norris, J. E., \& Cannon, R. D.
2006, \apj, 647, 1075

\bibitem[Stetson(1987)]{PS87} Stetson, P. B. 1987, \pasp, 99, 191

\bibitem[Stetson(1994)]{PS94} Stetson, P. B. 1994, \pasp, 106, 250

\bibitem[Zinn(1993)]{RZ93} Zinn, R. 1993, The Globular Cluster---Galaxy Connection (ASP
Conf.\ Ser.\ 48), ed.\ G. H. Smith \& J. P. Brodie (San Francisco, CA: ASP), 38

\bibitem[Zinn \& West(1984)]{ZW84} Zinn, R., \& West, M. J. 1984, \apjs, 55, 45

\end{thebibliography}
\end{document}